\documentclass [12pt]{article}
\usepackage {amssymb}
\usepackage {amsmath}
\usepackage[dvips]{epsfig}

\begin{document}

\begin{center}
\Large\textbf{Set of equations for transient enhanced diffusion in
shallow ion-implanted layers}
\\[2ex]
\normalsize
\end{center}

\begin{center}
\textbf{O. I. Velichko}$^{1 \ast}$, \textbf{Yu. P. Shaman}$^{1}$,
\textbf{A. K. Fedotov}$^{2}$, \textbf{and  A. V. Masanik}$^{2}$
\end{center}

\begin{center}
\bigskip

$^{1}$ \textit{Belarusian State University of Informatics and
Radioelectronics, 6, P. Brovki Str., Minsk, 220013 Belarus}

$^{2}$ \textit{Belarusian State University, 4, Nezavisimosti
Avenue, Minsk, 220050 Belarus}
\bigskip

$^{\ast}$ \textit{Corresponding author: E-mail:
oleg\_velichko@lycos.com; oleg\_velichko@yahoo.com}

\end{center}

\textit{Abstract.} To simulate the transient enhanced diffusion
near the surface or interface, a set of equations describing the
impurity diffusion and quasichemical reactions of dopant atoms and
point defects in shallow ion-implanted layers is proposed and
analyzed. The diffusion equations obtained take into account
different charge states of mobile or immobile species and drift
the mobile species in the built-in electric field and field of
elastic stresses. The absorption of self-interstitials on the
surface and drift of the defects due to elastic stresses result in
the nonuniform distributions of point defects. It was shown
analytically and by means of numerical calculations that
consideration of the nonuniform defect distributions enables one
to explain the phenomenon of ``uphill'' impurity diffusion near
the surface during annealing of ion-implanted layers. The
performed calculations of the boron concentration profile after
annealing of a shallow implanted layer agree well with the
experimental data confirming the efficiency of the proposed
equations.

\bigskip



\section{Introduction}
Using low-energy high-fluence ion implantation with the following
rapid thermal annealing, one can produce the active regions of
silicon devices characterized by very shallow junctions ($\sim $
0.1 $\mu $m) and high dopant concentrations
\cite{Solmi_03,Girginoudi_04,Boucard_05}. During annealing the
transient enhanced diffusion (TED) occurs. Therefore, the final
dopant distribution is determined both by ion implantation and
transient enhanced diffusion parameters. As the lateral dimensions
of integrated circuits are scaled down to the submicrometer range,
the need for accurate modeling of silicon doping is increased. The
well-known and widely used models of TED during rapid thermal
annealing of semiconductor substrates very often lead to the
results which disagree with the experimental data for low-energy
high-dose ion implantation, especially in the near surface region.
For example, the calculations presented in Refs.
\cite{Solmi_03,Girginoudi_04} provide some evidence of
disagreement with the experimental data in the vicinity of the
interface. The difference is mainly due to the use of inadequate
clustering models for the description of high concentration
diffusion \cite{Komarov_06}, due to the influence of interfaces on
the defect distributions \cite{Lamrani_04}, due to dopant atom
trapping by immobile sinks \cite{Ferri_06}, and due to the
influence of stresses on the dopant and defect diffusion
\cite{Lim_00,Aziz_01}. The principal goal of this investigation is
to obtain a set of equations describing impurity diffusion and
quasichemical reactions of dopant atoms and point defects near the
surface or interface to simulate adequately the transient enhanced
diffusion in shallow implanted layers.

\section{System of equations}
It is proposed that during the initial stage of annealing the
ion-implanted silicon substrates a thin damaged layer is formed
near the surface or in the vicinity of the interface. For example,
such a damaged layer can be formed due to incomplete solid phase
recrystallization of silicon regions heavily doped by As
\cite{Velichko_88} or due to the near surface absorption of mobile
point defects created by boron implantation. During annealing this
damaged layer can absorb mobile point defects and impurity atoms,
that results in the formation of nonuniform defect distributions
and electrical deactivation of impurity. Moreover, if the damaged
layer contains a great amount of defects and impurity atoms,
significant stresses can arise in the region adjoined to the
surface. To describe diffusion of point defects and impurity atoms
in this intricate system, the following set of equations is
proposed:

{\bf 1.} Expression for the total concentration of impurity atoms
$C^{T}$

\begin{equation}\label{Total}
 C^{T} = C + C^{AC} + C^{AD} \quad {\rm ,}
\end{equation}

\noindent where $C$ is the concentration of substitutionally
dissolved impurity atoms, $C^{AC}$ and $C^{AD}$ are the
concentrations of impurity atoms incorporated into clusters and
bound to the extended defects, respectively;

{\bf 2.} Conservation law for impurity atoms incorporated into
clusters (precipitates)

\begin{equation}\label{Cluster}
\frac{{\partial \,C^{AC}}}{{\partial \,t}} = S^{AC} - G^{AC} \quad
{\rm .}
\end{equation}

{\bf 3.} Conservation law for impurity atoms bound to the extended
defects

\begin{equation}\label{Extended}
\frac{{\partial \,C^{AD}}}{{\partial \,t}} = S^{AD} - G^{AD} \quad
{\rm .}
\end{equation}

{\bf 4.} Equation of diffusion for impurity atoms due to the
formation, migration, and dissociation of the pairs ``impurity
atom --- point defect'' which are in equilibrium with the
substitutionally dissolved impurity atoms and point defects
\cite{Velichko_97}. For the two-stream diffusion govern by the
vacancies and self-interstitials this equation has the form

\begin{equation}\label{Diffusion}
\begin{array}{l}
 \displaystyle{\frac{{\partial \,C}}{{\partial \,t}}} = {\left[ {D^{E}\;{\frac{{\partial
\left( {a^{E}\tilde {C}^{V\times} C} \right)}}{{\partial x}}} +
{\frac{{D^{E}a^{E}\tilde {C}^{V\times} C}}{{\chi} }}\;\;{\frac{{\partial
\chi} }{{\partial x}}}} \right]} - {\frac{{\partial} }{{\partial x}}}\left(
{v_{x}^{E} a^{E}\tilde {C}^{V\times} C} \right) \\
 \\ \quad \quad +  \displaystyle{{\frac{{\partial} }{{\partial x}}}}{\left[
{D\;^{F}{\frac{{\partial \left( {a^{F}\tilde {C}^{I\times} C}
\right)}}{{\partial x}}} + {\frac{{D^{F}a^{F}\tilde {C}^{I\times}
C}}{{\chi
}}}\;\;{\frac{{\partial \chi} }{{\partial x}}}} \right]} \\
 \\ \displaystyle{\quad \quad - {\frac{{\partial} }{{\partial x}}}}\left( {v_{x}^{F}
a^{F}\tilde {C}^{I\times} C} \right)\; - S^{A} + G^{A} \\
 \end{array}{\rm .}
\end{equation}

5. Equation of vacancy diffusion \cite{Velichko_03}

\begin{equation}\label{Vacancy}
 \begin{array}{l}
  \displaystyle{\frac{{\partial \;}}{{\partial \;x}}}{\left[ {d^{VC}(\chi
){\frac{{\partial \;(a^{Vd}\tilde {C}^{V\times} )}}{{\partial
\;x}}}} \right]} - {\frac{{\partial \;}}{{\partial \;x}}}\left(
{{\rm \tilde
{v}}_{{\rm x}}^{{\rm V}} \tilde {C}^{V\times} } \right) - \\
 \\-  \displaystyle{{\frac{{k^{VC}}}{{\left( {l_{i}^{V}}  \right)^{\;2}}}}\tilde {C}^{V\times
}} + {\frac{{\tilde {g}^{V}}}{{\left( {l_{i}^{V}}  \right)^{\;2}}}} = 0 \\
 \end{array} \quad {\rm .}
\end{equation}

 6. Equation for diffusion of self-interstitials \cite{Velichko_03}

\begin{equation}\label{Self-interstitial}
 \begin{array}{l}
  \displaystyle{\frac{{\partial \;}}{{\partial \;x}}}{\left[ {d^{IC}(\chi
){\frac{{\partial \;(a^{Id}\tilde {C}^{I\times} )}}{{\partial
\;x}}}} \right]} - {\frac{{\partial \;}}{{\partial \;x}}}\left(
{{\rm \tilde
{v}}_{{\rm x}}^{{\rm I}} \tilde {C}^{I\times} } \right) - \\
\\ -  \displaystyle{\frac{{k^{IC}}}{{\left( {l_{i}^{I}}
\right)^{\;2}}}}\tilde {C}^{I\times
} + {\frac{{\tilde {g}^{I}}}{{\left( {l_{i}^{I}}  \right)^{\;2}}}} = 0 \\
 \end{array} \quad {\rm ,}
\end{equation}

where the effective coefficients of the equations are presented as
follows:

\begin{equation}\label{Diffusivity_by_V}
 D^{E}\left( {\chi}  \right) = D_{i}^{E} {\frac{{1 + \beta _{1}^{E} \chi +
\beta _{2}^{E} \chi ^{2}}}{{1 + \beta _{1}^{E} + \beta _{2}^{E}}
}} \quad {\rm ,}
\end{equation}

\begin{equation}\label{Diffusivity_by_I}
 D^{F}\left( {\chi}  \right) = D_{i}^{F} {\frac{{1 + \beta _{1}^{F} \chi +
\beta _{2}^{F} \chi ^{2}}}{{1 + \beta _{1}^{F} + \beta _{2}^{F}}
}} \quad {\rm ,}
\end{equation}

\begin{equation}\label{Chi}
 \chi = {\frac{{C + { \displaystyle{\frac{{z^{Cl}}}{{m}}}}C^{AC} - C^{B} + \sqrt {\left( {C +
 \displaystyle{{\frac{{z^{Cl}}}{{m}}}}C^{AC} - C^{B}} \right)^{\,2} + 4n_{i}^{2}}
}}{{2n_{i}} }} \quad {\rm ,}
\end{equation}

\begin{equation}\label{Relative_VI}
 \tilde {C}^{V\times}  = {\frac{{C^{V\times} }}{{C_{eq}^{V\times} } }} \quad
{\rm ,} \quad \tilde {C}^{I\times}  = {\frac{{C^{I\times}
}}{{C_{eq}^{I\times} } }} \quad {\rm ,}
\end{equation}

\begin{equation}\label{Relative_Diffusivity_absorption}
 d^{VC,IC}\left( {\chi}  \right) = {\frac{{d^{V,I}\left( {\chi}
\right)}}{{d_{i}^{V,I}} }} \quad {\rm ,} \quad k^{VC,IC} =
k^{V,I}\;\tau _{i}^{V,I} \quad {\rm ,}
\end{equation}

\begin{equation}\label{Relative_migration_length}
 l_{i}^{V} = \sqrt {d_{i}^{V} \tau \;_{i}^{V}}  \quad {\rm ,}
\quad l_{i}^{I} = \sqrt {d_{i}^{I} \tau \;_{i}^{I}}  \quad {\rm ,}
\end{equation}

\begin{equation}\label{Relative_drift_velocity}
 {\rm \tilde {v}}_{{\rm x}}^{{\rm V}{\rm ,}{\rm I}} = {\frac{{{\rm v}_{{\rm
x}}^{{\rm V}{\rm ,}{\rm I}}} }{{d_{i}^{V,I}} }} \quad {\rm ,}
\quad \tilde {g}^{{\rm V}{\rm ,}{\rm I}} = {\frac{{g^{{\rm V}{\rm
,}{\rm I}}}}{{d_{i}^{V,I} C_{eq}^{V\times ,I\times} } }} \quad
{\rm .}
\end{equation}

Here $S^{AC}$ and $G^{AC}$ are the rates of impurity atom
absorption due to the cluster formation and generation of separate
impurity atoms during cluster dissolution, respectively; $S^{AD}$
and $G^{AD}$  are the rates of impurity atom absorption due to the
formation of extended defects and generation of separate impurity
atoms during extended defect annealing, respectively; $\tilde
{C}^{V\times}$ and $\tilde {C}^{I\times }$ are the concentrations
of vacancies and self-interstitials in the neutral charge state
normalized to the equilibrium concentrations $C_{eq}^{V\times} $
and $C_{eq}^{I\times}$, respectively; $D^{E}\left( {\chi} \right)$
is the effective diffusivity of impurity atoms due to the
vacancy---impurity pairs mechanism; $D^{F}\left( {\chi} \right)$
is the effective diffusivity of impurity atoms due to migration of
the pairs ``impurity atom --- self-interstitials'' $D_{i}^{E} $
and $D_{i}^{F}$ are the intrinsic diffusivities; superscripts $E$
and $F$ denote the vacancy and interstitial pair diffusion
mechanisms, respectively; $\beta {\kern 1pt} _{1}^{E,F} $  and
$\beta {\kern 1pt} _{2}^{E,F} $ are the parameters describing
respectively the relative contributions of singly and doubly
charged point defects to the impurity transport; functions
$a^{E,F}$  describe the overall influence of heavy doping on the
local equilibrium between diffusing pairs, point defects and
substitutionally dissolved dopant atoms and also between the point
defects in different charge states; $v_{x}^{E} $ and $v_{x}^{F} $
are the   projections onto the $x$ -axis of the effective drift
velocities of dopant atoms in the field of elastic stresses; $\chi
$  is the concentration of charge carriers normalized to the
intrinsic carrier concentration $n_{i} $; $C^{B}$ is the
concentration of impurity with opposite type of conductivity;
$z^{Cl}$ and $m$ are the cluster charge in units of the elementary
charge and the number of doping atoms incorporated in the cluster;
$d^{V,I}\left( {\chi}  \right)$ and $d_{i}^{V,I} $ are
respectively the effective and intrinsic diffusivities of point
defects; superscripts $V$  and $I$  denote the vacancies and
self-interstitials, respectively; $k^{V,I}$ and $\tau _{i}^{V,I} $
are the   effective recombination coefficient and the average
lifetime of the corresponding point defects, respectively;
functions $a^{Vd,Id}$ describe the influence of heavy doping on
the local equilibrium between point defects in different charge
states; $v_{x}^{V} $ and $v_{x}^{I} $ are the projections onto the
$x$  -axis of the effective drift velocities of vacancies and
self-interstitials in the field of elastic stresses, respectively;
$l_{i}^{V} $  and $l_{i}^{I} $  are the average migration lengths
of vacancies and self-interstitials, respectively; $g_{i}^{V} $
and $g_{i}^{I}$  are respectively the generation rates of
vacancies and self-interstitials.

As follows from Ref. \cite{Komarov_06}, to describe the impurity
clustering during transient enhanced diffusion of As and P, one
can use the expression

\begin{equation}\label{Cluster_concentration}
 C^{AC} = K\tilde {C}_{D} \chi ^{m -
z^{Cl}}C^{m}
\end{equation}

\noindent instead of Eq.(\ref{Cluster}), if the annealing duration
is not very short (for example, annealing is longer than 1 s at a
temperature of 1000 $^{\circ}$C and higher). A similar expression
can be used for the description of phosphorus clustering
\cite{Velichko_TBP}.

Here $K$ is the constant of the cluster formation reaction; $m$ is
the number of impurity atoms incorporated in cluster; $\tilde
{C}_{D} $  is the relative concentration of nonequilibrium defects
participating in clustering

\begin{equation}\label{Cluster_defect}
 \tilde {C}_{D} = {\frac{{(\tilde {C}_{D1} )^{m_{1}} }}{{(\tilde {C}_{D2}
)^{m_{2}} }}} \quad {\rm ,} \quad \tilde {C}_{D1} =
{\frac{{C^{D1\times} }}{{C_{eq}^{D1\times} } }} \quad {\rm ,}
\quad \tilde {C}_{D2} = {\frac{{C^{D2\times} }}{{C_{eq}^{D2\times}
} }} \quad {\rm ,}
\end{equation}

\noindent  where $C^{D1\times} $  and $C^{D2\times} $  are the
concentrations of point defects ${\rm D}_{{\rm 1}}^{\times}  $ and
${\rm D}_{{\rm 2}}^{\times}$  in the neutral charge states
facilitating the cluster formation and generated during
clustering, respectively; $C_{eq}^{D1\times} $ and
$C_{eq}^{D2\times} $  are the equilibrium concentrations of these
defects; $m_{1} $  and $m_{2} $  are respectively the numbers of
point defects ${\rm D}_{{\rm 1}} $  and ${\rm D}_{{\rm 2}} $
participating in the cluster formation. Due to multiplier $\tilde
{C}_{D}$  in the expression (\ref{Cluster_concentration}), it is
possible to take into account the influence exerted by the
nonuniform distributions of nonequilibrium point defects on the
clustering process and influence of the defect generation during
clustering on transient enhanced diffusion. Thus, the expression
(\ref{Cluster_concentration}) enables one to consider the coupled
phenomenon of clustering and transient enhanced diffusion.

\section{Analysis of equations}

  The set of Eqs.(\ref{Total}) --- (\ref{Self-interstitial}) can describe different
processes of diffusion and chemical reactions in semiconductors
including ``uphill'' impurity diffusion in the vicinity of the
interface. The proposed equations take into account different
charge states of all mobile and immobile species and also the
drift of the mobile species in the built-in electric field and in
the field of elastic stresses, although only the concentrations of
neutral defects are given in the explicit form in these equations.
  The phenomenon of ``uphill'' diffusion can arise due
to the direct absorption of impurity atoms by extended defects
\cite{Ferri_06} or due to clustering in the near surface region,
if these processes are nonequilibrium and proceed in the forward
direction. To describe this phenomenon one can use
Eqs.(\ref{Cluster}) and (\ref{Extended}), respectively. Moreover,
as can be seen from Eq.(\ref{Diffusion}), the ``uphill'' impurity
diffusion occurs due to the influence of stresses on the pair
migration \cite{Velichko_97} or due to the formation of nonuniform
distributions of point defects in the neutral charge state
\cite{Velichko_84}. As it follows from Eqs.(\ref{Vacancy}) and
(\ref{Self-interstitial}), the nonuniform distributions of
vacancies and self-interstitials can be formed due to the
absorption of point defects by the surface (interface) or by
extended defects. Moreover, the nonuniform defects distributions
can be formed due to the stress-mediated migration of vacancies
and self-interstitials. Thus, a multifactor character of the
``uphill'' impurity diffusion in the near surface region follows
from analysis of Eqs.(\ref{Total}) --- (\ref{Self-interstitial}),
and it is difficult to determine which factor is dominant in the
formation of the near surface peak of the impurity concentration.
However, there is an experimental evidence that the silicon
surface absorbs self-interstitials \cite{Lamrani_04}, which govern
the transient enhanced diffusion. It means that in any case the
formation of nonuniform distributions of point defects due to
their absorption on the surface provides a certain contribution to
the ``uphill'' impurity diffusion.

Using an analytical approach, let us evaluate the influence of the
nonuniform neutral defect distribution on the ``uphill''
diffusion. Consider the case of TED governed by one species of
nonequilibrium point defects, namely nonequilibrium
self-interstitials. Let us suppose for simplicity that the
concentration of an impurity providing the opposite type of
conductivity is approximating zero $C^{B} \approx 0$ and $a^{F}
\approx 1$. Also, suppose that during annealing the formation of
neutral clusters or precipitates occurs. Then, in the region of
high impurity concentration Eq.(\ref{Diffusion}) takes the form

\begin{equation}\label{Diffusion_High}
{\frac{{\partial C^{T}}}{{\partial \,t}}} = {\frac{{\partial}
}{{\partial x}}}\left( {2D^{F}\left( {\chi}  \right)\;\tilde
{C}^{I\times }{\frac{{\partial C}}{{\partial x}}}\; +
D\;^{F}\left( {\chi} \right)\;C{\frac{{\partial \tilde
{C}^{I\times} }}{{\partial x}}}\;} \right) \quad {\rm .}
\end{equation}

The first term in the right-hand side parentheses describes the
impurity flux according to the Fick's law, whereas the second one
describes an additional flux arising due to the nonuniform defect
distribution. If this flux is opposite to that due to the impurity
concentration gradient, the phenomenon of ``uphill'' diffusion is
observed.

For simplicity, we consider the case of long-time annealing. Then,
a steady-state solution of Eq. (16) for $t \to \infty $  and
$\displaystyle{\frac{{\partial C^{T}}}{{\partial \,t}}} \approx 0$
can be derived

\begin{equation}\label{Steady_State_Solution}
 2D^{F}\left( {\chi}  \right)\;\tilde {C}^{I\times} {\frac{{\partial
C}}{{\partial x}}}\; + D^{F}\left( {\chi}
\right)\;\;C{\frac{{\partial \tilde {C}^{I\times} }}{{\partial
x}}} = A = const \quad {\rm .}
\end{equation}

Let us consider the structure ``silicon-on-insulator'' and suppose
reflecting boundary conditions for impurity atoms on the surface
($x = 0$) and at the interface ($x = x_{R} $). Then $A = 0$, and
Eq. (17) can be written as

\begin{equation}\label{Steady_State_Reflecting}
 {\frac{{dC}}{{dx}}} = - b\left( {x} \right)\,C \quad {\rm ,}
\end{equation}

\noindent  where

\begin{equation}\label{Space_Function}
 b\left( {x} \right) = {\frac{{1}}{{2\tilde {C}^{\times} }}}{\frac{{d\tilde
{C}^{I\times} }}{{dx}}} \quad {\rm .}
\end{equation}

We can solve Eq.(\ref{Steady_State_Solution}) using the method of
variable separation

\begin{equation}\label{Variable_Separation}
{\frac{{dC}}{{C}}} = - b\left( {x} \right)\,dx \quad {\rm ,} \quad
{\int\limits_{C_{R}} ^{C} {{\frac{{dC}}{{C}}}}}  = -
{\int\limits_{x_{R} }^{x} {b\left( {x} \right)\,dx}}  \quad {\rm
,}
\end{equation}

\begin{equation}\label{Solution}
C\left( {x} \right) = C_{R} \exp {\left[ {{\int\limits_{x}^{x_{R}}
{b\left( {x} \right)\,dx}} } \right]} \quad {\rm ,}
\end{equation}

\noindent  where $C_{R} = C\left( {x_{R}}  \right)$ is the
impurity concentration at the interface.

Let us suppose for simplicity that the distribution of
self-interstitials is approximately described by the solution of
Eq.(\ref{Self-interstitial}) with Dirichlet boundary conditions
and constant equation coefficients taking the influence of
stresses and generation of point defects in the near surface
region as negligible

\begin{equation}\label{Defect_Solution}
\tilde {C}^{I\times} \left( {x} \right) = \sinh ^{ - 1}\left(
{{\frac{{x_{R} }}{{l_{i}^{I}} }}} \right){\left[ {\tilde
{C}_{S}^{I\times}  \sinh \left( {{\frac{{x_{R} - x}}{{l_{i}^{I}}
}}} \right) + \tilde {C}_{R}^{I\times} \sinh \left(
{{\frac{{x}}{{l_{i}^{I}} }}} \right)} \right]} \quad {\rm ,}
\end{equation}

\noindent where $\tilde {C}_{S}^{I\times}$ and $\tilde
{C}_{R}^{I\times}$ are the normalized concentrations of point
defects at the surface and interface, respectively.

Substituting expression (\ref{Space_Function})  into
(\ref{Solution}), one can obtain

\begin{equation}\label{Integral_Solution}
C\left( {x} \right) = C_{R} \exp {\left[
{{\frac{{1}}{{2}}}{\int\limits_{\tilde {C}^{I\times} }^{\tilde
{C}_{R}^{I\times} }  {{\frac{{d\tilde {C}^{I\times} }}{{\tilde
{C}^{I\times }}}}\,}} } \right]} \quad {\rm .}
\end{equation}

Integral calculation in expression (\ref{Integral_Solution}) gives

\begin{equation}\label{Final_Solution}
C\left( {x} \right) = C_{R} \exp {\left[ {{\frac{{1}}{{2}}}\ln
\left( {\tilde {C}_{R}^{I\times} }  \right) - {\frac{{1}}{{2}}}\ln
\left( {\tilde {C}^{I\times} } \right)} \right]} = C_{R} \sqrt
{{\frac{{\tilde {C}_{R}^{I\times} } }{{\tilde {C}^{I\times} \left(
{x} \right)}}}} \quad {\rm .}
\end{equation}

It can be seen from expression (\ref{Final_Solution}) that the
steady-state distribution of impurity atoms varies inversely as
the square root from the normalized self-interstitial distribution
$\tilde {C}^{I\times} \left( {x} \right)$. It means that for the
defect distribution, similar to the distribution of (22) with
$\tilde {C}_{S}^{I\times}  < < \tilde {C}_{R}^{I\times} $  , i.e.
in case of the absorption of self-interstitials at the surface,
the ``uphill'' impurity diffusion and peak of impurity
concentration adjacent to the surface must be observed. It follows
from Eqs. (\ref{Total}) and (\ref{Cluster_concentration}) that
this peak will be more pronounced in the case of significant
clustering of impurity atoms.

\bigskip

\section{Simulation results}

\begin{figure}[ht]
\centering
\includegraphics[width=5.50in,height=3.85in]{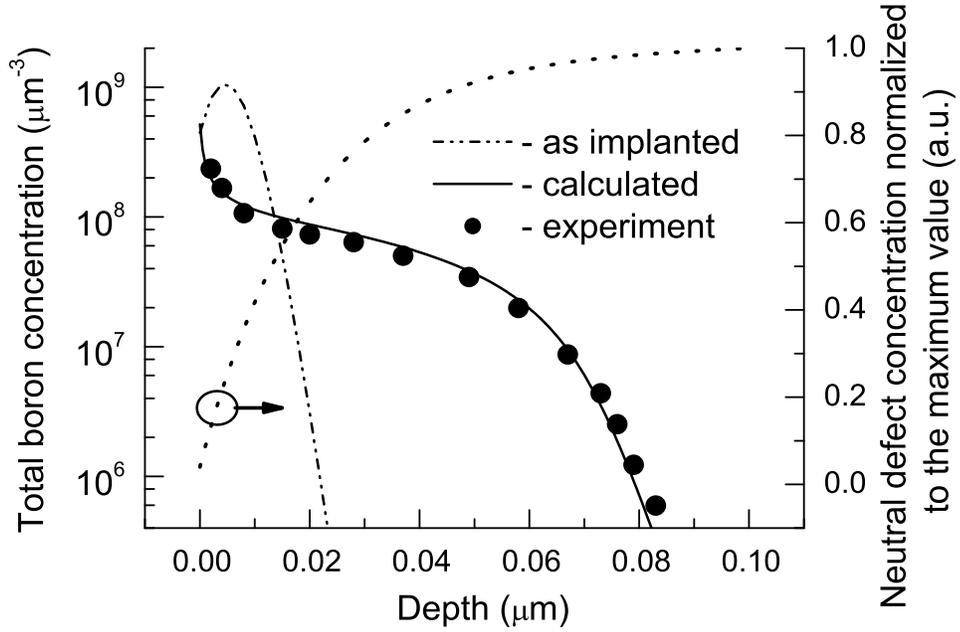}
\caption{Calculated concentration profile of ion-implanted boron
after thermal annealing during 10 s at a temperature of 1050
$^{\circ}$C. Solid line --- calculated distribution of the total
boron concentration after annealing; dash-dotted line ---
distribution of boron after ion implantation; dotted line --- time
average distribution of self-interstitial concentration. Circles
represent the experimental data taken from Lerch et al.
\cite{Lerch_99}. \label{fig:Boron}}
\end{figure}

Results of the previous analytical investigation are confirmed by
the numerical simulation presented below. As an example, the
redistribution of boron atoms implanted into silicon at low energy
and high fluence is investigated. For comparison, the experimental
data obtained by Lerch et al. \cite{Lerch_99} are used. In Ref.
\cite{Lerch_99} boron ions were implanted into n-type silicon
substrates of the resistivity 10-20 $\Omega $m and orientation
(100) at the energy of 1 keV and fluence $Q$ = 1.0$\times
$10$^{{\rm 1}{\rm 5}}$ cm$^{{\rm -} {\rm 2}}$. The annealing
process was carried out at a temperature of 1050 $^{\circ}$C with
duration of 10 s. The concentration profile of the boron atom
distribution was measured by SIMS. This concentration profile is
presented in Fig.~\ref{fig:Boron} together with the profile
calculated by means of the set of equations obtained.

The finite-difference method \cite{Samarskii_01} is used to find a
numerical solution of this system of equations. The present paper
chooses a more stable implicit method to avoid numerical
instability. Following \cite{Samarskii_01}, the right-hand side of
Eq.(\ref{Diffusion}) and two first terms on the left-hand side of
Eq.(\ref{Self-interstitial}) which describe the diffusion of
dopant atoms and self-interstitials, respectively, are
approximated by a symmetric difference operator of second order
accuracy on the space variable $x$. To solve the system of the
obtained nonlinear algebraic equations we use the simplest
iterative technique substituting the values of the dopant
concentration and concentration of neutral self-interstitials
determined on the previous iteration into the coefficients of
nonlinear algebraic equations.

During simulation it was supposed that transient enhanced
diffusion of the boron implanted into silicon substrate occurred
due to the kick-out diffusion mechanism that is mathematically
equivalent to the pair diffusion mechanism \cite{Robinson_92}. We
have used the following parameters to describe ion implantation
and diffusion of the dopant atoms and point defects: $Q$ =
1.0$\times $10$^{{\rm 1}{\rm 5}}$ cm$^{{\rm -} {\rm 2}}$; $R_{p} $
= 0.00512 $\mu$; $\Delta R_{p} $ = 0.0038 $\mu $; $Sk$ = 0.3;
$n_{i} $ = 1.166$\times $10$^{{\rm 7}} \; \mu ^{{\rm - }{\rm 3}}$;
$D_{i} \tilde {C}_{\max} ^{I\times}  $ = 1.3$\times $10$^{{\rm -
}{\rm 0}{\rm 5}} \; \mu ^{{\rm 2}}$/s; $\beta _{1}^{I}$ = 0.7;
$\beta _{2}^{I}$ = 0; $\tilde {C}_{S\max} ^{I\times}  =
{{C_{S}^{I\times} } \mathord{\left/ {\vphantom {{C_{S}^{I\times} }
{C_{\max} ^{I\times} } }} \right. \kern-\nulldelimiterspace}
{C_{\max} ^{I\times} } }$ =0.04 a.u.; $l_{i}^{I} $ = 0.02 $\mu$.
Here $R_{p} $ is the average projected range of dopant ions;
$\Delta R_{p}$ is the straggling of the projected range; $Sk$ is
the skewness of boron distribution as implanted; $D_{i}$ is the
intrinsic diffusion coefficient of boron; $\tilde {C}_{\max}
^{I\times}  = {{C_{\max} ^{I\times} } \mathord{\left/ {\vphantom
{{C_{\max} ^{I\times} } {C_{eq}^{I\times} } }} \right.
\kern-\nulldelimiterspace} {C_{eq}^{I\times} }}$ is a maximum
value of the normalized self-interstitial concentration; $C_{\max}
^{I\times}$ is a maximum value of the neutral self-interstitial
concentration in the region of impurity diffusion;
$C_{S}^{I\times}  $ is the concentration of neutral
self-interstitials at the surface.
  The value of equilibrium boron diffusivity at 1050
$^{\circ}$C known from literature $D_{i}$ = 4.098$\times
$10$^{{\rm -} {\rm 6}} \; \mu ^{{\rm 2}}$/s \cite{Haddara_00}.
This means that the time average enhancement of boron diffusion
due to the nonequilibrium self-interstitials is approximately 3.2
fold.

As can be seen from Fig.~\ref{fig:Boron}, the measured boron
profile is characterized by the formation of a local peak of the
impurity concentration near the surface due to the distinct
``uphill'' diffusion. It is interesting to note that the impurity
profiles calculated by means of the computer code SSUPREM-IV
included in the desktop program (framework) Athena which are
presented in \cite{Lerch_99} do not describe this phenomenon. In
our opinion, this discrepancy with the experiment in the near
surface region is due to neglecting the interface effect on the
defect subsystem that is absolutely inappropriate for the models
intended to simulate shallow doped layers. It was shown in
\cite{Lamrani_04} that the semiconductor surface absorbs
nonequilibrium self-interstitials governing the transient enhanced
diffusion of the ion-implanted impurity. As can be seen from
expression (\ref{Final_Solution}) and numerical calculations
presented in Fig.~\ref{fig:Boron}, taking into account the
phenomenon of self-interstitial absorption enables one to explain
the formation of a near surface peak of the impurity concentration
and provides a good agreement of the calculated impurity profile
with the experimental data. It is interesting to note that
absorption of boron atoms by extended defects in the near surface
region also can contribute to the ``uphill'' diffusion. In this
case the nonuniform self-interstitial distribution with decreased
concentration gradient can be used to explain the experimental
data.

\newpage

\section{Conclusion}

For adequate simulation of the transient enhanced diffusion near
the surface or at the interface of ion-implanted layers the set of
equations describing   impurity diffusion and quasichemical
reactions of the dopant atoms and defects was proposed and
analyzed. The proposed diffusion equations take into account
different charge states of mobile and immobile species and drift
of the mobile species in the built-in electric field and the field
of elastic stresses, although in these equations only the
concentrations of neutral defects are given in the explicit form.
The absorption of self-interstitials on the surface and drift of
mobile defects due to elastic stresses result in the nonuniform
distributions of point defects. It was shown, analytically and by
means of numerical calculations, that consideration of the
nonuniform defect distributions makes it possible  to explain the
phenomenon of ``uphill'' impurity diffusion near the surface
during annealing of the ion-implanted layers. The performed
calculations of the boron concentration profile after annealing of
a shallow implanted layer agree well with the experimental data
including the near surface region, where the formation of a
concentration peak occurs. The agreement with experiment confirms
the efficiency of the proposed equations for simulation of shallow
layer annealing.

\end{document}